\documentclass[aps,prd,superscriptaddress,floatfix,showpacs,nofootinbib]{revtex4}

\usepackage{verbatim}

\usepackage{amsmath}
\usepackage{amsfonts}
\usepackage{amssymb}
\usepackage{graphicx}
\usepackage{bm}
\usepackage{color}
\usepackage{epsf}

\usepackage{psfrag}

\def\bea{\begin{eqnarray}}
\def\eea{\end{eqnarray}}
\def\beas{\begin{eqnarray*}}
\def\eeas{\end{eqnarray*}}
\def\beqas{\begin{eqnarray*}}
\def\eqas{\end{eqnarray*}}
\def\beq{\begin{equation}}
\def\eeq{\end{equation}}
\def\beqd{\begin{displaymath}}
\def\eeqd{\end{displaymath}}
\def\eqd{\end{displaymath}}

\def\slashchar#1{\setbox0=\hbox{$#1$}
   \dimen0=\wd0
   \setbox1=\hbox{/} \dimen1=\wd1
   \ifdim\dimen0>\dimen1
      \rlap{\hbox to \dimen0{\hfil/\hfil}}
      #1
   \else\begin{eqnarray}
      \rlap{\hbox to \dimen1{\hfil$#1$\hfil}}
      /
   \fi}

\begin{document}
\title
{Electroproduction of a large invariant mass photon pair}
\author{A.~Pedrak}
\affiliation{National Centre for Nuclear Research (NCBJ), Pasteura 7, 02-093 Warsaw, Poland}
\author{ B.~Pire}
\affiliation{ Centre de Physique Th\'eorique, CNRS, \'Ecole Polytechnique, I. P. Paris,
  91128 Palaiseau,     France  }
\author{ L.~Szymanowski}
\affiliation{National Centre for Nuclear Research (NCBJ), Pasteura 7, 02-093 Warsaw, Poland}
\author{ J.~Wagner}
\affiliation{National Centre for Nuclear Research (NCBJ), Pasteura 7, 02-093 Warsaw, Poland}
\begin{abstract}
We study the exclusive electroproduction of a photon pair in the kinematical regime where the diphoton invariant mass is large and where  the nucleon flies almost in its original direction. We discuss the relative importance of the QCD process where the two photons originate from a quark line compared to the single (double) Bethe-Heitler processes where one (two) photons originate from the lepton line. This process turns out to be a promising tool to study generalized parton distributions in the nucleon, both at the medium energy of JLab and at a high energy electron ion collider.\end{abstract}
\maketitle
\section{Introduction}
 Hard electroproduction processes are a powerful probe of the hadronic structure. Their understanding in the framework of the QCD collinear factorization  of hard amplitudes in specific kinematics in terms of generalized parton distributions (GPDs) and hard perturbatively calculable coefficient functions \cite{historyofDVCS, gpdrev} offers a way to access the inner structure of nucleons and light nuclei.

\begin{figure}
\includegraphics[width=10cm]{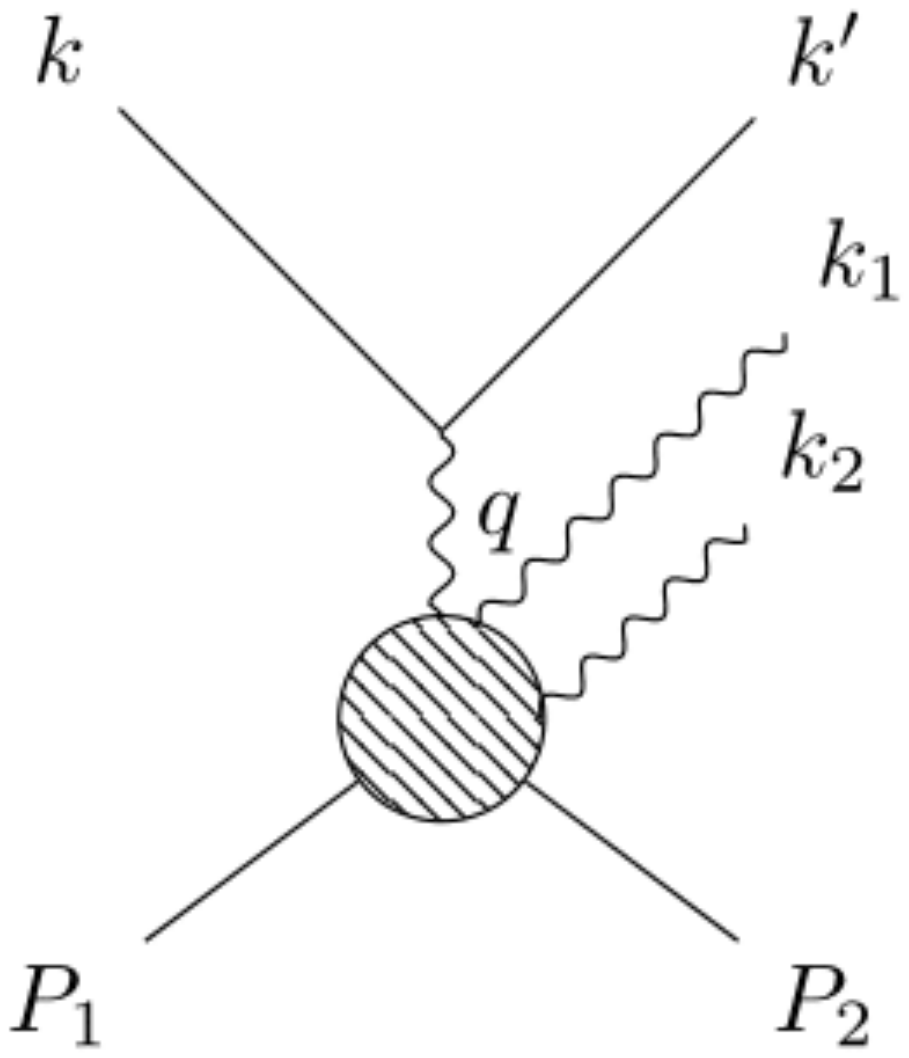}
\vspace{-4cm}
\caption{The QCD process contributing to $e N \to e' \gamma \gamma N'$. }
\label{0BH}
\end{figure}

\begin{figure}
\includegraphics[width=16cm]{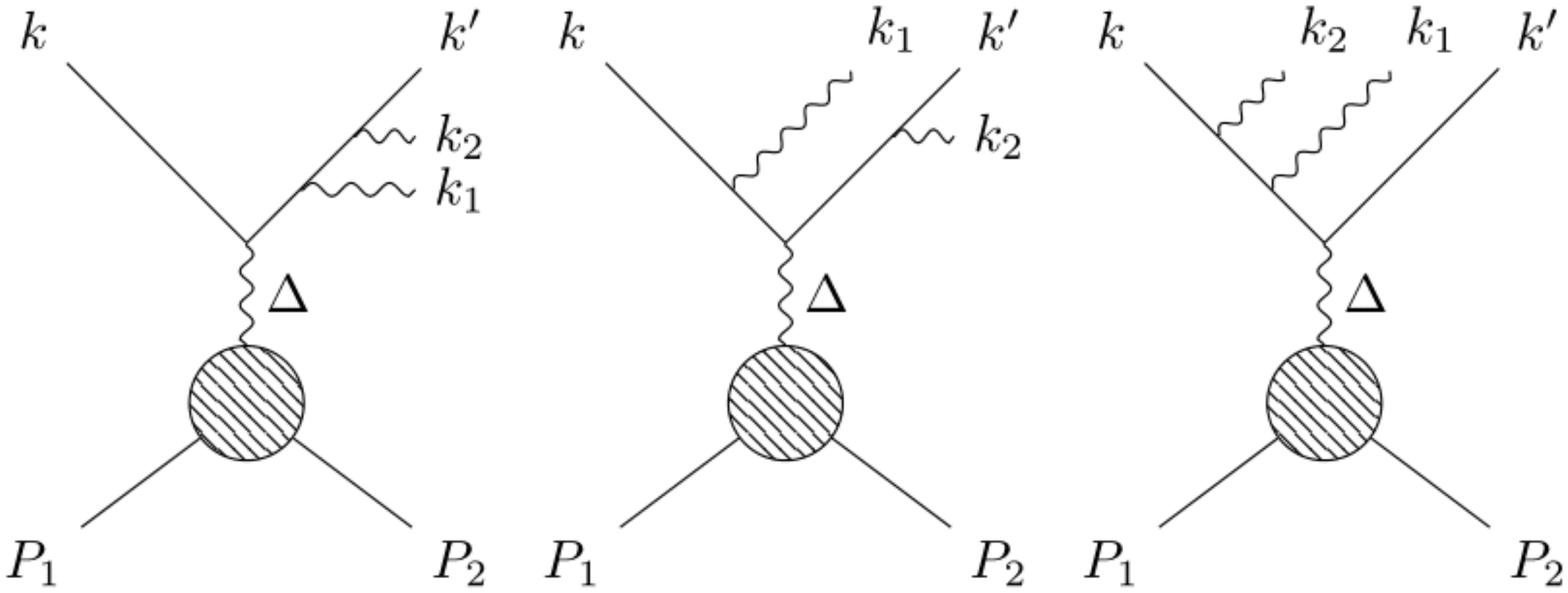}
\vspace{-3cm}
\caption{The double Bethe-Heitler process contributing to $e N \to e' \gamma \gamma N'$. Three other graphs with $k_1 \leftrightarrow k_2$ interchange are not shown.}
\label{2BH}
\end{figure}

In a previous  work \cite{Pedrak:2017cpp, Pedrak:2019vcf}, we studied the exclusive photoproduction of two photons on an   unpolarized proton or neutron target
\begin{equation}
\gamma(q,\epsilon) + N(P_1,s_1) \rightarrow \gamma(k_1,\epsilon_1) +  \gamma(k_2,\epsilon_2)+ N'(P_2,s_2)\,,
\label{processgamma}
\end{equation}
 in the kinematical regime of large invariant diphoton squared mass  $M_{\gamma\gamma}^2=(k_1+k_2)^2$ of the final photon pair and small momentum transfer $t =(P_2-P_1)^2$ between the initial and the final nucleons. This process shares many features with timelike Compton scattering (TCS), the photoproduction of a large mass lepton pair \cite{TCS}. We enlarge our study to the more general case of electroproduction
\begin{equation}
e(k,\lambda) + N(P_1,s_1) \rightarrow e'(k',\lambda) + \gamma(k_1,\epsilon_1) +  \gamma(k_2,\epsilon_2)+ N'(P_2,s_2)\,,
\label{process}
\end{equation} 
which generalizes the real photoproduction process (\ref{processgamma}) to the virtual photon case
\begin{equation}
\gamma^*(q,\epsilon) + N(P_1,s_1) \rightarrow \gamma(k_1,\epsilon_1) +  \gamma(k_2,\epsilon_2)+ N'(P_2,s_2)\,,
\label{processgammastar}
\end{equation}
with $q=k-k'$ and $q^2\neq 0$. 
 As will be demonstrated below, this reaction has a number of interesting features, both for moderate energy in the JLab domain and for the high energy domain of a future electron-ion collider (EIC)~\cite{Accardi:2012qut}, and can be used as an  important source of information for future programs aiming at the extraction of GPDs~\cite{PARTONS}. There are three processes contributing to the reaction (\ref{process}), namely the QCD process of Fig.~\ref{0BH} where the two photons are emitted from a quark,  the single Bethe-Heitler process of Fig.~\ref{1BH} where one photon is emitted from a lepton and the other photon from a quark, and the double Bethe-Heitler process of Fig.~\ref{2BH} where the two photons originate from a lepton. As demonstrated below, their relative importance depends very much on the kinematical conditions, and particularly on the value of $Q^2=-(k'-k)^2$.

 The plan of this paper is the following. In section II, we review the kinematics of the reaction. Section III presents the calculation of the amplitudes of the three processes which contribute and shows the main features of their contributions to the differential cross-sections. In section IV, we compare their relative contributions to the differential cross section from medium (JLab) to high (EIC) energies and  examine to which extent the quasi-real electroproduction of a large mass diphoton can be described by the photoproduction cross-section with the known flux of Weizs{\"a}cker-Williams equivalent photons. Section V  gathers some conclusions. Appendix presents the spinor techniques used in our calculations. 
\section{Kinematics}
\label{Sec:Kinematics}
Let us first present the kinematics of the process (\ref{process}).
 It is most similar to the one of double deeply virtual Compton scattering (DDVCS) \cite{DDVCS}. 
We decompose every momenta on a Sudakov basis  as
\begin{equation}
\label{sudakov1}
v^\mu = a \, n^\mu + b \, p^\mu + v_\bot^\mu \,,
\end{equation}
with $p$ and $n$ the light-cone vectors
\begin{equation}
\label{sudakov2}
p^\mu = \frac{\sqrt{s}}{2}(1,0,0,1)\,,\qquad n^\mu = \frac{\sqrt{s}}{2}(1,0,0,-1)\,, \qquad p\cdot n = \frac{s}{2}\,,
\end{equation}
and
\begin{equation}
\label{sudakov3}
 v^+ = v \cdot n  \,, \qquad v_\bot^\mu = (0,v^x,v^y,0) \,, \qquad v_\bot^2 = -\vec{v}_t^2\,.
\end{equation}

The particle momenta, in the chosen reference frame, read
\begin{eqnarray}\label{impfinc}
k^\mu &=& \lvert \vec k \rvert (1,\sin \theta \cos\phi,\sin \theta \sin\phi, \cos \theta)=\frac{\lvert \vec k \rvert}{\sqrt s}[(1+\cos\theta)p^\mu +(1-\cos \theta)n^\mu]+k_\bot^\mu \, ,\nonumber \\
q^\mu &=& y n^\mu -\frac{Q^2}{ys} p^\mu ~,\quad k'^\mu= k^\mu - q^\mu \,,\nonumber \\
 P_1^\mu &=& (1+\xi)\,p^\mu + \frac{M^2}{s(1+\xi)}\,n^\mu~, \quad P_2^\mu = (1-\xi)\,p^\mu + \frac{M^2+\vec{\Delta}^2_t}{s(1-\xi)}n^\mu + \Delta^\mu_\bot\,, \nonumber \\
k_1^\mu &=& \alpha_1 \, n^\mu + \frac{(\vec{p}_t-\vec\Delta_t/2)^2}{\alpha_1 s}\,p^\mu + p_\bot^\mu -\frac{\Delta^\mu_\bot}{2}~,~~
 k_2^\mu = \alpha_2 \, n^\mu + \frac{(\vec{p}_t+\vec\Delta_t/2)^2}{\alpha_2 s}\,p^\mu - p_\bot^\mu-\frac{\Delta^\mu_\bot}{2}\,,
\end{eqnarray}
where 
$M$ is the mass of the nucleon and $\xi$ is the skewness parameter.  We define the momentum transfer $\Delta^\mu = P_2^\mu-P_1^\mu$. In the $\gamma \gamma$ center of mass system, we define the angles $\theta_{\gamma \gamma}$, $\phi_{\gamma \gamma}$ as the polar and azimuthal angles of $\gamma(k_1)$.
Since the azymuthal dependence of the process can only depend  on the difference $\phi_{\gamma \gamma} -\phi$, we fix by convention $ \phi =0$.

The total center-of-mass energy squared of the electron-nucleon system $S_{e N}$ is, neglecting terms of order $M^2/S_{e N}$ or $t/S_{e N}$:
\begin{equation}
\label{energysquared}
S_{e N} \approx (1+\xi) s\,,
\end{equation}
and the skewness parameter equals
\begin{equation}
\label{xidef}
\xi\approx \frac{Q^2+M_{\gamma\gamma}^2}{2yS_{eN}-Q^2-M_{\gamma\gamma}^2}\,.
\end{equation}
Since we here enlarge our previous study of the photoproduction process, we define the  large (factorization) scale as 
\begin{eqnarray}
\label{Mgg}
M_{\gamma\gamma}^2 &=& ~(k_1 +k_2)^2 = \frac{4 \vec p_t ^2}{\sin ^2\theta_{\gamma\gamma}}\,, 
\end{eqnarray}
which, as usual, is quite arbitrary  but sufficient for a leading order computation; $t=\Delta ^2$ is kept small with respect to this scale. 

We will write the differential cross-section of the process as
\begin{equation}
\label{cs}
\frac{d\sigma}{dQ^2dydtd\phi dM_{\gamma\gamma}^2d\Omega_{\gamma\gamma}} = \frac{1}{2}\frac{\alpha_{em}^4}{16(2\pi)^3} \frac{1}{(S_{eN}-M^2)^2} \lvert {\cal M} \rvert ^2\,.
\end{equation}
Momentum conservation puts a lower value to $y$ since $y S_{eN}$ must be larger than  $M_{\gamma\gamma}^2$. For JLab energies, this $y_{min}$ is not small.
\section{The production processes}

We choose to calculate, at leading order in $\alpha_{em}$ and $\alpha_S$,  the amplitudes of the three contributing processes following the Kleiss-Sterling spinor techniques \cite{Kleiss:1985yh}, with the external photon polarization vectors in the gauge defined by the $p$ vector. Useful formulas are gathered in Appendix. This is to our knowledge the first time that this formalism is used for a hard exclusive process. We neglect everywhere the lepton mass.

\subsection{The QCD process}
 \begin{figure}
\includegraphics[width=14cm]{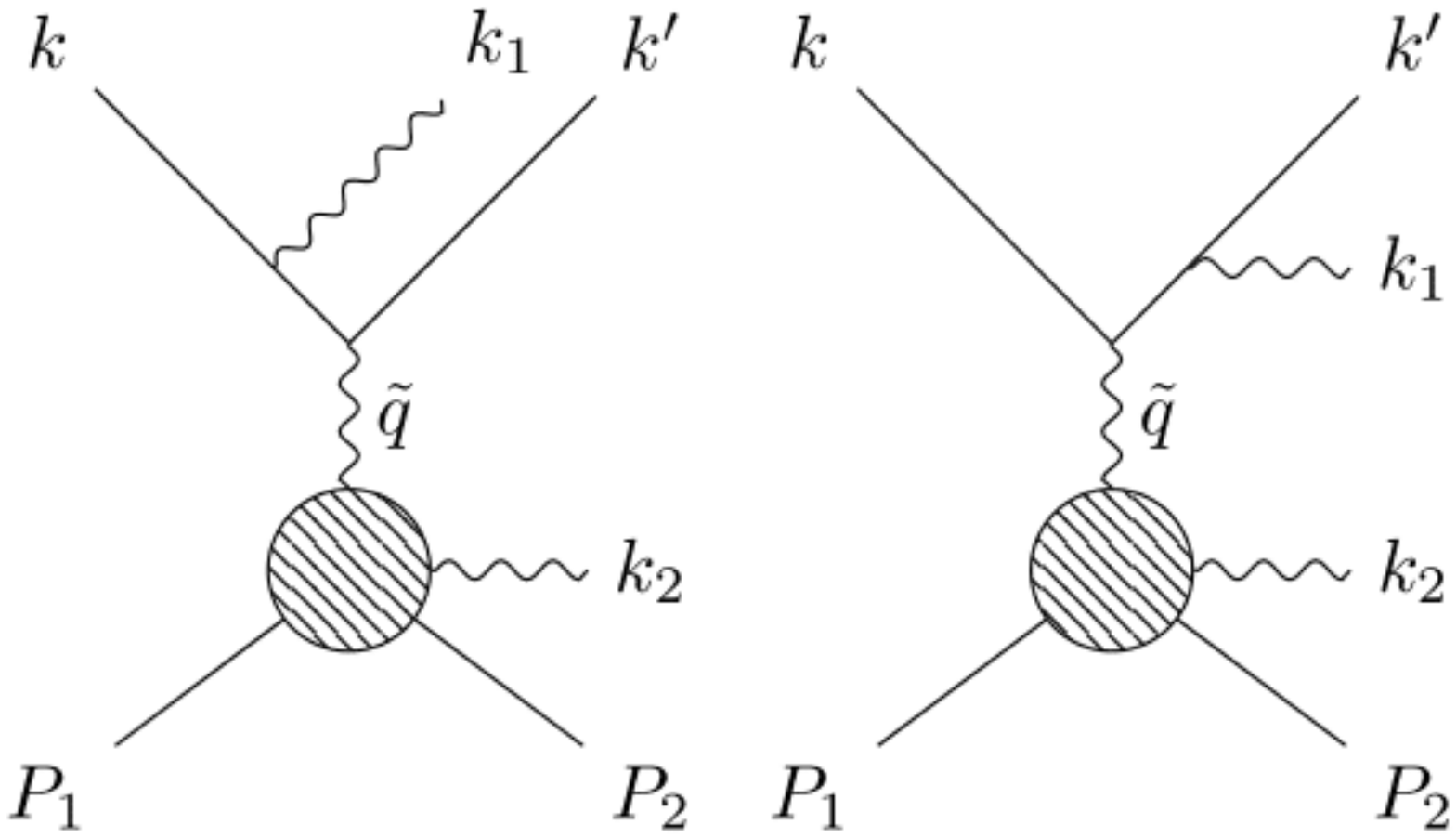}
\vspace{-2cm}
\caption{The single Bethe-Heitler process contributing to $e N \to e' \gamma \gamma N'$. Two other graphs with $k_1 \leftrightarrow k_2$ interchange are not shown.}
\label{1BH}
\end{figure}

\begin{figure}
\includegraphics[width=12cm]{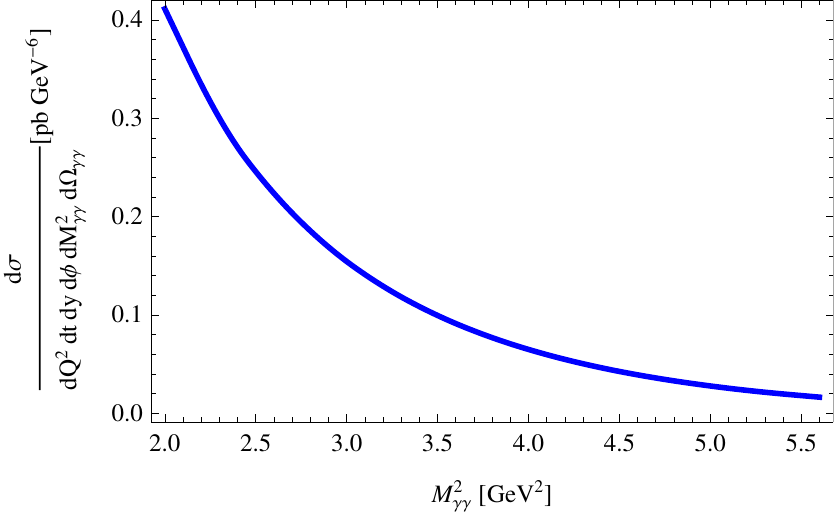}
\caption{
 The $M_{\gamma\gamma}^2-$ dependence of the QCD process contribution to the  $e N \to e' \gamma \gamma N'$ differential cross section, at $Q^2=10^{-3}$~GeV$^2$, $s=20$~GeV$^2$, $y=0.8$, $\theta=3\pi/8$, $\phi_{\gamma\gamma} = 0$.
}
\label{0BH_PsiDep}
\label{0BH_mggDep}
\end{figure}

\begin{figure}
\includegraphics[width=8cm]{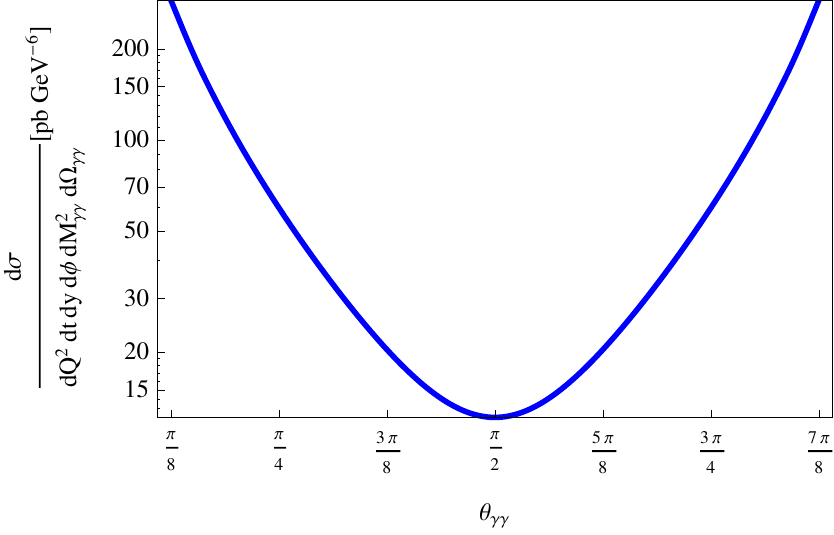}~~~~~~~\includegraphics[width=8cm]{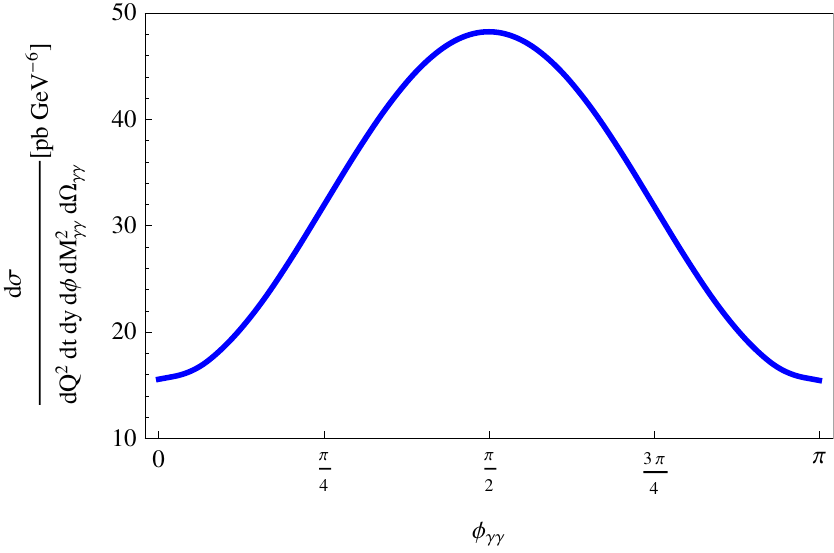}
\caption{The angular dependence (in the $\gamma \gamma$ center of mass system) of the QCD process contribution to the  $e N \to e' \gamma \gamma N'$ cross section,  at $s= 20$ GeV$^2$, $y=0.6$,$M_{\gamma\gamma}^2 = 3$ GeV$^2$ and $Q^2=10^{-5}$ GeV$^2$, as a function of $\theta_{\gamma\gamma}$ for $\phi_{\gamma\gamma}=\pi/8$ (left panel) and as a function of $\phi_{\gamma\gamma}$ for $\theta_{\gamma\gamma}=3\pi/8$ (right panel).}
\label{AzDep}
\end{figure}
Calculating the amplitude of the QCD process follows the same lines as in the photoproduction case \cite{Pedrak:2017cpp}. We detail in Appendix the case of the helicity amplitude ($+ + \to + + - +$). Other cases are similar.

The simplified kinematical relations used to calculate the hard coefficient functions are :
\begin{eqnarray}
\label{simplkin}
&&\alpha_1 + \alpha_2 = y~~~~,~~~~\alpha_1  = \frac{y}{2}\left[1+\sqrt{1-\frac{4\vec{p}_t^2}{2\xi s y -Q^2}}\,\right] \,, \nonumber\\
&& \lvert \vec k \rvert = \frac{\sqrt s}{2} \left[1+\frac{Q^2}{s}\frac{1-y}{y^2} \right]~~~~,~~~~\cos \theta = - \frac{1-\frac{Q^2}{s}\frac{1-y}{y^2}}{1+\frac{Q^2}{s}\frac{1-y}{y^2}} \,~~~.
\end{eqnarray}

Because of the charge conjugation properties of the ($\gamma^*, \gamma_1, \gamma_2$) system, the QCD process turns out to be sensitive only to the non-singlet combination of GPDs:
\begin{equation}
H^{(-)}(x)=\sum_q Q_q^3\left(H^q(x)+H^q(-x)\right)\,.
\end{equation}
There is no  contribution from the singlet quark, nor gluonic GPDs, at any order of the QCD calculation.

\begin{figure}
 \includegraphics[width=7cm]{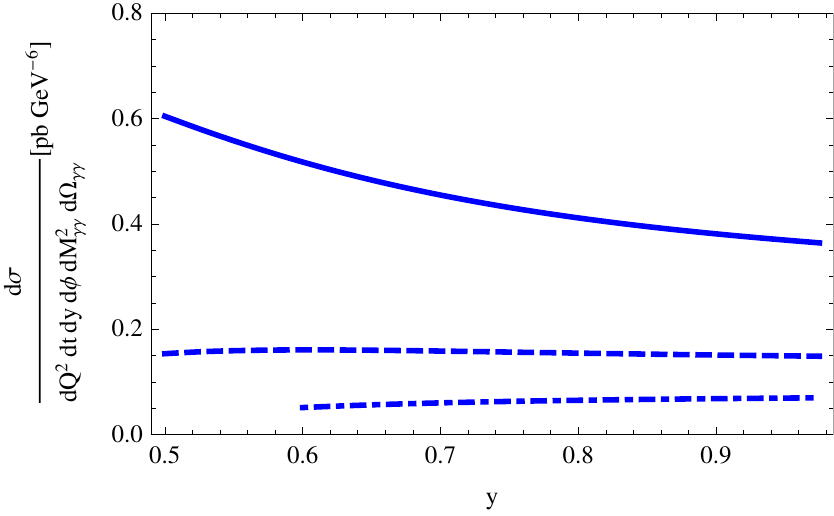}
\caption{The $y-$dependence of the QCD process contribution to the  $e N \to e' \gamma \gamma N'$ differential cross section, at $Q^2=10^{-3}$~GeV$^2$, $s=20$~GeV$^2$ and $M_{\gamma\gamma}^2= 2$~GeV$^2$ (full curve),  $M_{\gamma\gamma}^2= 3$~GeV$^2$ (dashed curve) and $M_{\gamma\gamma}^2= 4$~GeV$^2$ (dash-dotted curve), at $\theta=3\pi/8$, $\phi_{\gamma\gamma} = 0$.}
\label{ydep0BH}
\end{figure}

Let us now present some features of the contribution of the QCD process to the unpolarized differential cross-section 
$\frac{d\sigma^{eN\to e'\gamma\gamma N'}}{d Q^2 dt dy d\phi dM^2_{\gamma\gamma}d\Omega _{\gamma\gamma}}$ at $t=t_{min}$ (i.e. for $\Delta_T=0$), $s=20$ GeV$^2$. For numerical estimates we used the Goloskokov-Kroll model \cite{GK} for GPD $H$ neglecting all other contributions (which were responsible for less then 1\% of the cross section in the photoproduction case). Since, for very small values of $Q^2$, this cross-section scales like $1/Q^2$, as it should following the Weizs{\"a}cker-Williams formula, the results for different small $Q^2$ values than presented on the plots can easily be deduced. The $M^2_{\gamma\gamma}$ dependence of the differential cross sections, shown in Fig.~\ref{0BH_mggDep} for $Q^2= 10^{-3}$ GeV$^2$, follows an effective powerlike behaviour $1/M^{2n}_{\gamma\gamma}$ with $n\approx 3$. We show on  Fig.~\ref{AzDep} the $\gamma \gamma $ angular dependence in the $\gamma \gamma $ center of mass system. The range in $\theta_{\gamma\gamma}$ is limited due to the requirement that the factorization scale for the single Bethe-Heitler process is high enough - as discussed in the section \ref{sec:1BH}. 
The  $y-$dependence shown in Fig.\ref{ydep0BH} is quite weak (provided $y>y_{min}$). The  magnitude of the cross section indicates that the discussed process is accessible in the currently running and designed experiments at JLAB and EIC.

\subsection{The single Bethe-Heitler process}
\label{sec:1BH}
Some care is needed to apply the collinear factorization framework to calculate the QCD part of the single Bethe-Heitler amplitude of Fig.~\ref{1BH}.  Indeed, the photon entering the hadronic part must be hard enough to justify a QCD description of the virtual Compton scattering sub-process $\gamma^*(\tilde q)N \to \gamma N'$. Since
\begin{equation}
    \tilde q = q - k_1 ~~~~~ \mathrm{or} ~~~~~ \tilde q = q - k_2 \,,
\end{equation}
one should distinguish the case where $-\tilde q ^2$ is large enough to apply QCD factorization, and the converse case where a description in terms of nucleon polarizabilities \cite{Fonvieille:2019eyf} is more appropriate. In practice, we shall only consider the kinematics which prevent $-\tilde q ^2$ from getting smaller than $1$ GeV$^2$, which is assured by choosing diphoton squared mass above $3$ GeV$^2$.

One should define properly the kinematics so that the factorized formalism can be straightforwardly applied. This implies to define a new Sudakov basis where the $\gamma^*(\tilde q) N(P_1)$ define the $(+, -)$ basis. We choose to do so with
\begin{eqnarray}
    \tilde p^\mu &=& p^\mu ~~;~~~ \tilde n^\mu = n^\mu - \frac{p_\bot ^2}{2np}p^\mu + p_\bot^\mu ~~~ \rm{when}  ~~~ \tilde q = q - k_1 \,,\\
    \tilde p^\mu &=& p^\mu ~~;~~~ \tilde n^\mu = n^\mu - \frac{p_\bot ^2}{2np}p^\mu - p_\bot^\mu ~~~ \rm{when}  ~~~ \tilde q = q - k_2 \,, 
\end{eqnarray}
so that $\tilde p \cdot \tilde n = p\cdot n$. The hard part is independent of these variations on the Sudakov vector $n$. Then we write the hadronic part of the single Bethe-Heitler amplitude with the replacement
\begin{equation}
\frac{H^q(x)}{s}\bar{U}(P_2,s_2)\not n U(P_1,s_1) \longrightarrow \frac{H^q(x)}{s}\bar{U}(P_2,s_2)\not \tilde n U(P_1,s_1)\,,
\label{nchoice}
    \end{equation}
for the contribution of the $H^q(x,\xi,t) $ GPD, and similar terms for the contributions of other GPDs\footnote{Let us remark that we encounter here a theoretical uncertainty  related to the choice of the light-like vector $n^\mu$ spanning the longitudinal subspace, which results in the appearance  of kinematical ambiguities of predictions usually attributed to higher twists effects. As discussed in  \cite{BraunManashov}, the predictions for a leading twist-2 contribution to the scattering amplitude are sensitive to the choice of $n^\mu$ appearing in the factorisation formula and in the parametrization of momenta. 
The full analysis of kinematical ambiguities in our process along Refs.\cite{BraunManashov} is obviously beyond the scope of our paper.}.

As the usual DVCS amplitude, because of the charge conjugation properties of the ($\gamma^*, \gamma$) system, the single Bethe-Heitler process turns out to be sensitive to the singlet combination of GPDs:
\begin{equation}
H^{(+)}(x)=\sum_q Q_q^2\left(H^q(x)-H^q(-x)\right)\,,
\end{equation}
 and (at higher order in $\alpha_s$) will benefit from gluonic  GPD contributions. NLO QCD corrections calculated for DVCS \cite{PSW} can  straightforwardly be applied to this process.

\begin{figure}
\includegraphics[width=10cm]{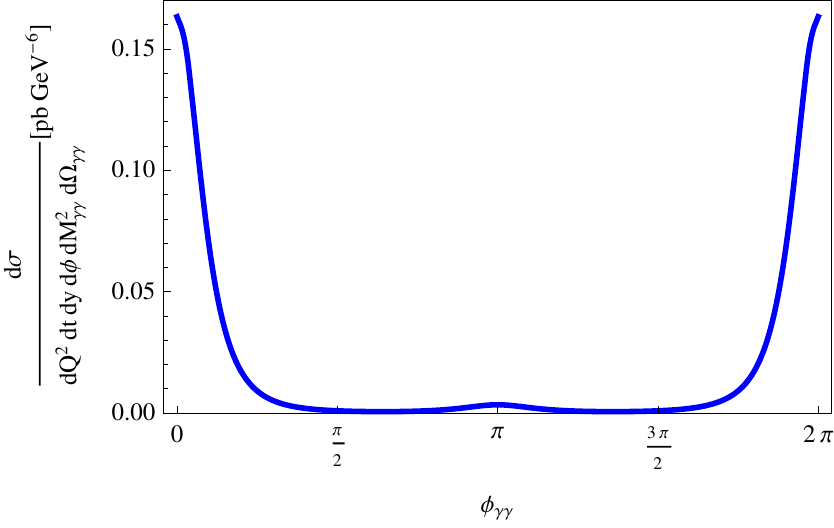}
\caption{The azymuthal dependence of the single Bethe-Heitler process contribution to the  $e N \to e' \gamma \gamma N'$ cross section, at $s=20$ GeV$^2$, $Q^2=1$ GeV$^2$, $y=0.6$, $\theta_{\gamma \gamma}=3\pi/8$ and $M_{\gamma\gamma}^2= 3$ GeV$^2$.}
\label{Phidep}
\end{figure}

\begin{figure}
 \includegraphics[width=7cm]{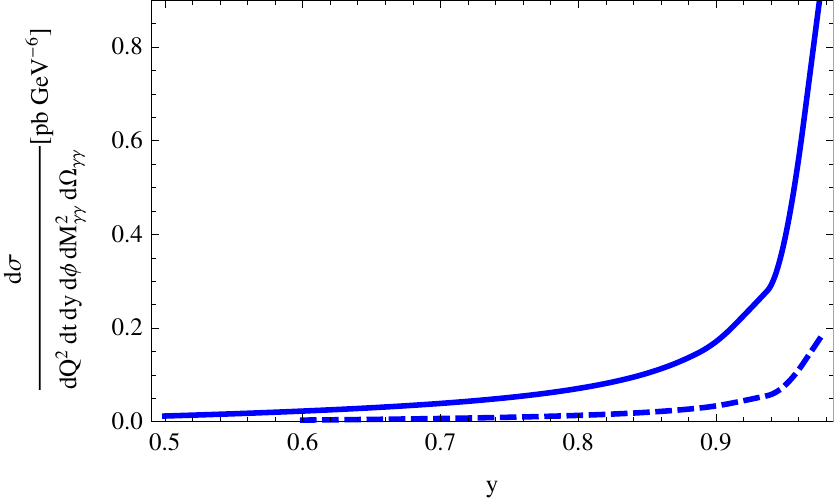}~~~~~~~~~\includegraphics[width=7cm]{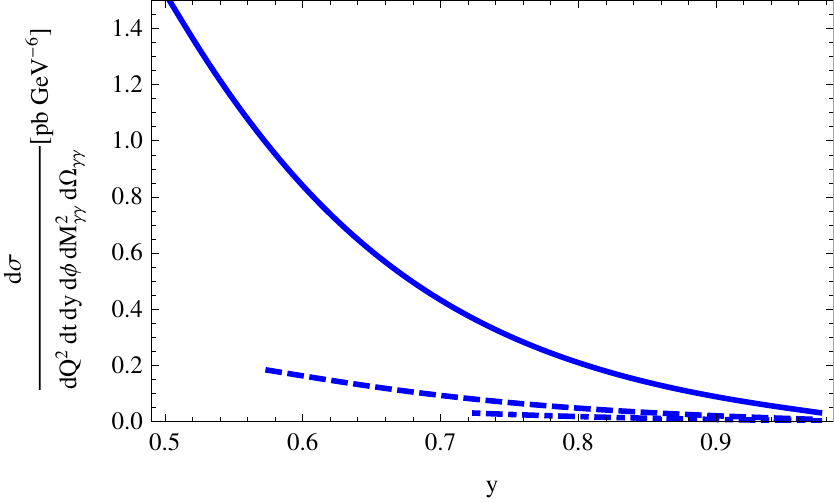}
\caption{The $y-$dependence of the single Bethe Heitler process contribution to the  $e N \to e' \gamma \gamma N'$ cross section, at $s=20$ GeV$^2$; Left panel : $Q^2=10^{-3}$ GeV$^2$ and $M_{\gamma\gamma}^2= 3$ GeV$^2$ (solid curve) and  $M_{\gamma\gamma}^2= 4$ GeV$^2$ (dashed curve); Right panel : $Q^2=1$ GeV$^2$ and $M_{\gamma\gamma}^2= 2$ GeV$^2$ (solid curve),  $M_{\gamma\gamma}^2= 3$ GeV$^2$ (dashed curve), $M_{\gamma\gamma}^2= 4$ GeV$^2$ (dash-dotted curve). }
\label{ydep}
\end{figure}

The $\phi_{\gamma \gamma}$ dependence of the single Bethe Heitler process contribution is plotted on Fig. \ref{Phidep}. We show on Fig. \ref{ydep} the  $y-$dependence of the single Bethe-Heitler process contribution to the cross-section, for both very small and sizeable $Q^2$ (there is no curve for the $M^2_{\gamma\gamma}=2$~GeV$^2$ on the left plot, as the factorization scale is too small in that case for a description in terms of GPD). 
 
\subsection{The double Bethe-Heitler process}
\begin{figure}[h!]
\includegraphics[width=7cm]{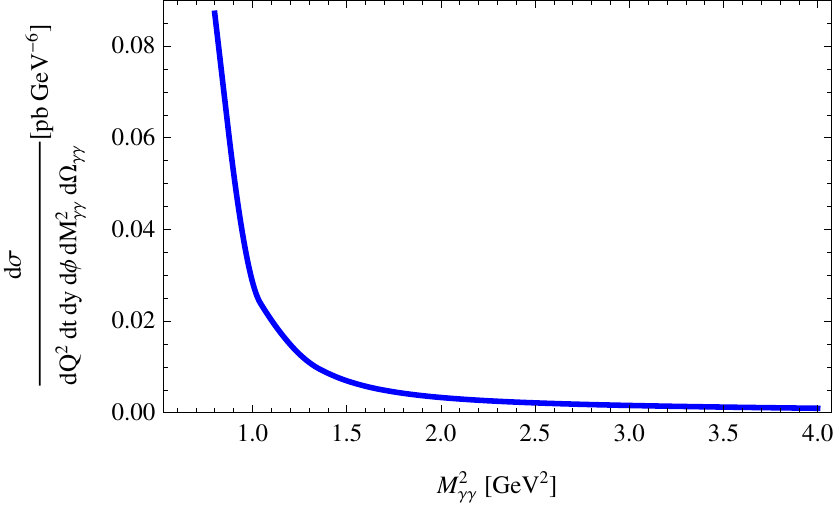}~~~~~~~~~\includegraphics[width=7cm]{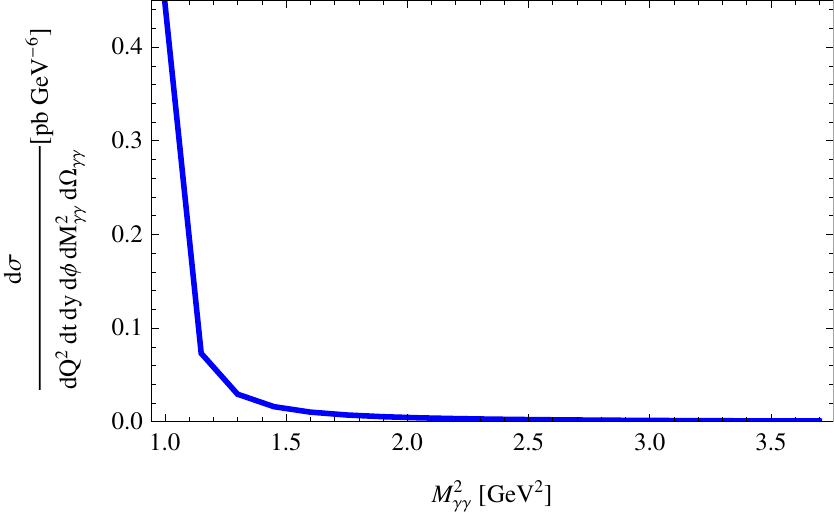}

\includegraphics[width=7cm]{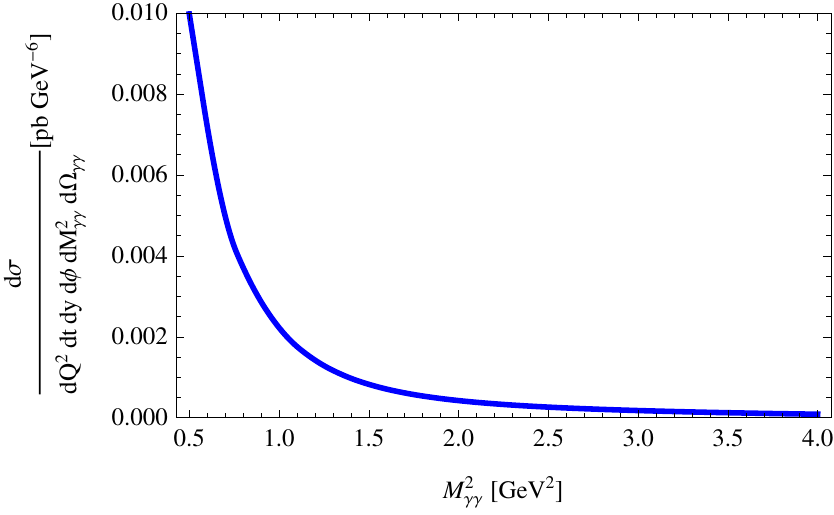}~~~~~~~\includegraphics[width=7cm]{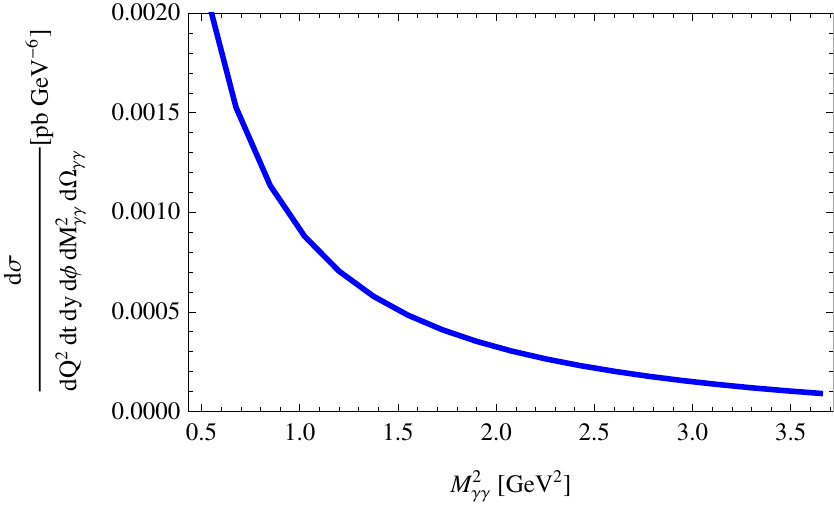} 
\caption{$M_{\gamma\gamma}^2$ dependence of the contribution of the double Bethe-Heitler process to the differential cross-section at $\theta=3\pi/8$, $y=0.8$, for $\phi_{\gamma\gamma} = 0$ (first line) or $\phi_{\gamma\gamma} = \pi/2$ (second line), and for $Q^2= 1$ (resp. $2$) GeV$^2$ on the left (resp. right) panel.   }
\label{2BHmgg}
\end{figure}
As the Bethe-Heitler contribution to the DVCS amplitude, the double Bethe-Heitler contribution is expressed in terms of the known  Dirac and Pauli electromagnetic form factors $F_1(\Delta^2)$ and $F_2(\Delta^2)$ as
\begin{equation}
    {\cal M}_{BH} = \frac{- 1}{\Delta^2} \bar u (k',\lambda){\cal L}^\mu u (k,\lambda) \bar U(P_2,s_2)\left[\gamma^\mu F_1(\Delta^2) + \sigma^{\mu\nu} \frac{i\Delta^\nu}{2M} F_2(\Delta^2)     \right] U(P_1,s_1)\,,
\end{equation}
with
\begin{eqnarray}
  {\cal L}^\mu &=&  \gamma^\mu \frac{1}{\hat k' +\hat \Delta +i \epsilon} \hat \epsilon_1 \frac{1}{\hat k' +\hat \Delta +\hat k_1 +i \epsilon}\hat \epsilon_2 + \hat \epsilon_1  \frac{1}{\hat k' +\hat k_1 +i \epsilon} \gamma^\mu \frac{1}{\hat k' +\hat \Delta +\hat k_1 +i \epsilon}\hat \epsilon_2    \nonumber \\
   &&  + \hat \epsilon_1  \frac{1}{\hat k' +\hat k_1 +i \epsilon} \hat \epsilon_2  \frac{1}{\hat k' +\hat k_2 +\hat k_1 +i \epsilon}\gamma^\mu   ~~~+~~~\left(k_1 \leftrightarrow k_2 \right) \,.
\end{eqnarray}
We calculate all the helicity amplitudes following the spinor techniques briefly described in the Appendix.

For illustration, we show on Fig. \ref{2BHmgg} the $M_{\gamma\gamma}^2$ dependence of the contribution of the double Bethe-Heitler process to the differential cross-section. Since our calculation is valid without any restriction on kinematics, we show the differential cross section even for rather small values of $M_{\gamma\gamma}^2$. This contribution decreases quite quickly with $M_{\gamma\gamma}$ and is more sizeable when $\phi_{\gamma\gamma}$ is small.
\section{Comparison of the processes}

\begin{figure}
\includegraphics[width=8cm]{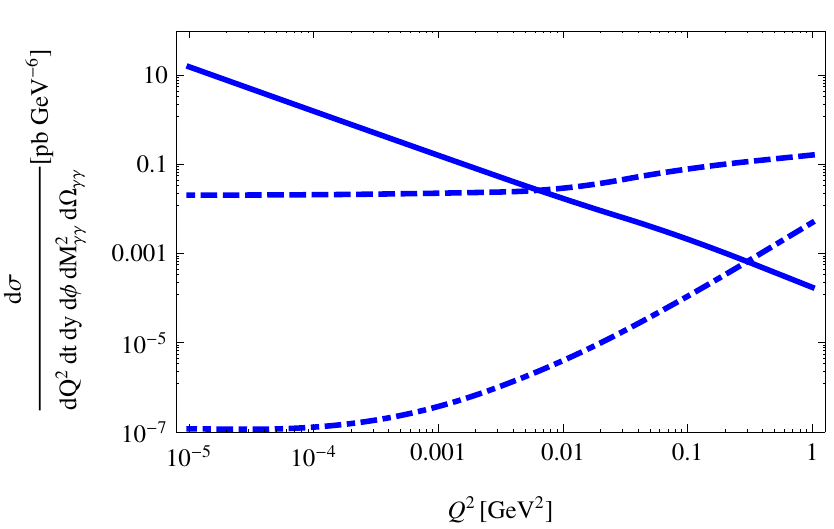}~~~~~\includegraphics[width=8cm]{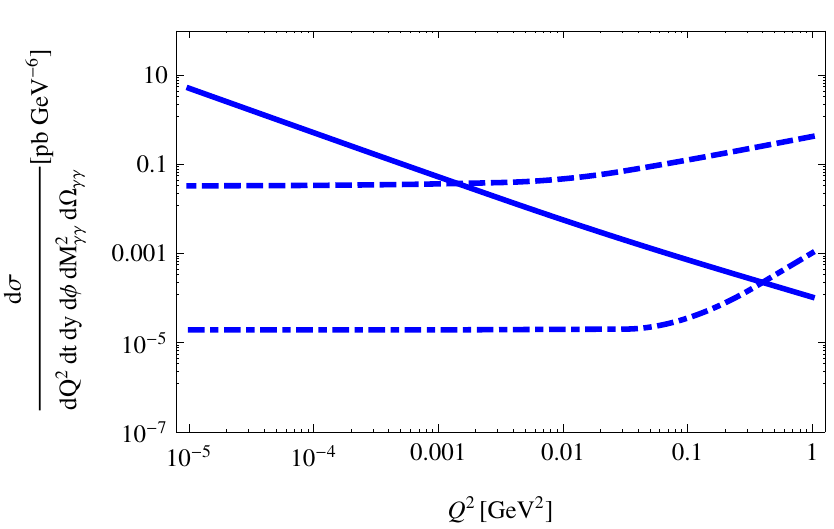}

\includegraphics[width=8cm]{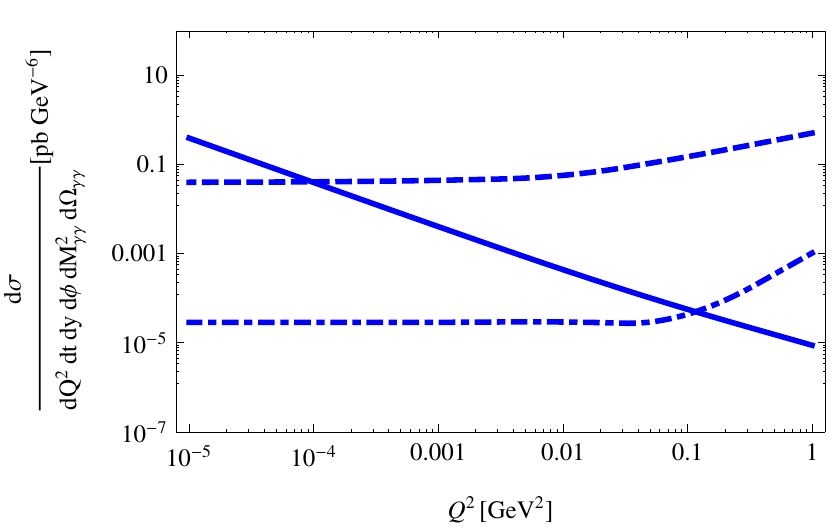}~~~~~\includegraphics[width=8cm]{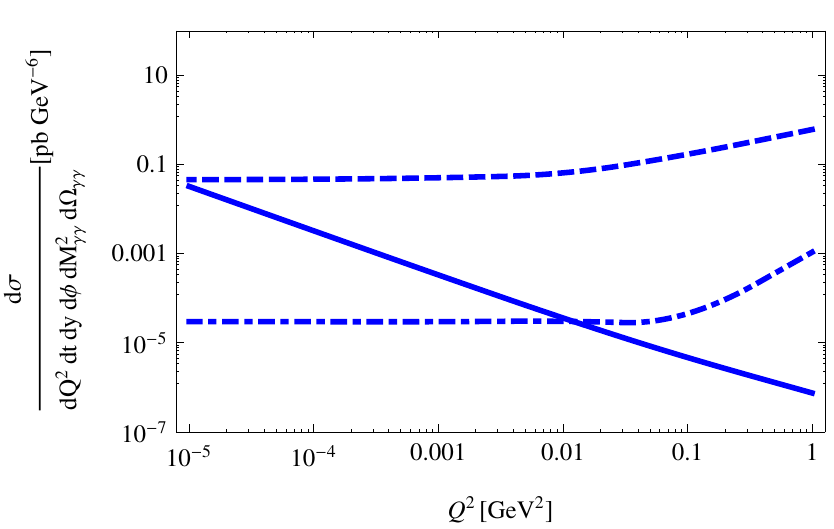}
\caption{The relative importance of the different processes contributing to $e N \to e' \gamma \gamma N'$ - shown here (from left to right and from top to bottom) for $s=20$GeV$^2$, $s=100$GeV$^2$, $s=1000$GeV$^2$ and $s=10000$GeV$^2$  at the kinematical point $M_{\gamma\gamma}^2 = 3$ GeV$^2$, $\theta=3\pi/8$, $\phi_{\gamma\gamma} = 0$, $y=0.6$ - depends much on the value of $Q^2$. The QCD process (solid curve) dominates at very low $Q^2$, the single Bethe-Heitler process (dashed curve) dominates at higher $Q^2$ while the double Bethe-Heitler process (dotted curve) is always subdominant. }
\label{Comp}
\end{figure}

The relative importance of the different processes contributing to $e N \to e' \gamma \gamma N'$ is illustrated in Fig. \ref{Comp}  where we plot the contributions of the three processes (neglecting their interferences) at a characteristic kinematical point. Their magnitude indeed depends much on the value of $Q^2$. The QCD process (solid curve) dominates at very low $Q^2$, the single Bethe-Heitler process (dashed curve) dominates at higher $Q^2$ while the double Bethe-Heitler process (dotted curve)  is always much smaller than either the QCD or the single Bethe-Heitler process in the  kinematical range we are  interested in; we can thus neglect this latter contribution for any phenomenological purpose.   

In the quasi-real photoproduction limit, we recover the results of our previous work  \cite{Pedrak:2017cpp} : the C-odd (valence) GPDs are accessed in a peculiar way, since the QCD amplitude is proportional to these GPDs taken at their border values $x=\pm \xi$. By scrutinizing the $Q^2$ dependence of the single Bethe-Heitler and the QCD amplitudes, we can  quantify what we mean by "quasi-real", which turns out to depend very much on the overall energy domain.  One may conclude that one can safely apply the  Weizs{\"a}cker-Williams equivalent photon approximation~\cite{Kessler:1975hh,Frixione:1993yw} for $Q^2< 10^{-3}$~GeV$^2$ at JLab but only for $Q^2< 10^{-5}$~GeV$^2$ at EIC.

\section{Conclusion}
Our calculation of the leading order leading twist amplitude of reaction (\ref{process})  has demonstrated that the electroproduction of a large invariant mass diphoton is an interesting process to analyze in the collinear factorization framework, both at current experimental facilities such as JLab and Compass at CERN, but also in future high energy experiments. The amplitude has very specific properties which should be very useful for future GPDs extractions programs e.g. \cite{PARTONS}.

One may sum-up our results as
\begin{itemize}
    \item On the one hand, the QCD process dominates the amplitude in the very small $Q^2$ domain for JLab energies. We quantified the maximal value of $Q^2$ for the Weizs{\"a}cker-Williams approximation to be valid. The conclusions written in our previous work on the photoproduction process then apply to an electroproduction experiment. For completeness, let us remind the reader that the main conclusion of this study is that the amplitude only depends of C-odd GPDs at the border values $x=\pm \xi$.
    \item On the other hand, the single Bethe-Heitler process strongly dominates the amplitude in the domain where $Q^2$ is not extremely small, especially in the large energy domain of the EIC.  We can then apply collinear QCD factorization to the amplitude provided that the virtuality of the exchanged photon is large enough and we verified that this was indeeed the case for large enough values of $M_{\gamma\gamma}^2$, typically above $3$ GeV$^2$.
In this window, the diphoton electroproduction coalesces to a deeply virtual Compton scattering but with the interesting difference that - because of the smallness of the double Bethe-Heitler amplitude - it is not anymore polluted by the Bethe-Heitler process which usually dominates the QCD process. This opens a new window for the extraction of C-even GPDs, and in particular for gluon GPDs which have been shown to give a large contribution in next to leading order (NLO)  $O(\alpha_S)$ calculations \cite{PSW}. This NLO study is however out of the scope of the present paper.
\end{itemize}

Although very important for the nucleon tomography program \cite{impact}, we did not detail the dependence on $\Delta_\bot$ which enters through the $t-$dependence of GPDs. Neither did we enter the discussion of higher twist effects, which should be taken into account as much as possible as in the DVCS case \cite{BraunManashov,twist3}. Finally, let us stress that NLO QCD corrections are likely to be important and should definitely be computed before a sensible phenomenology of our process is undertaken. Together with a further hint of the factorization of our process, it should yield an intersting information on the analytic structure of these QCD corrections in a process where both timelike $M_{\gamma\gamma}^2$ and spacelike $Q^2$ scales coexist, in contradistinction with the DVCS vs TCS cases  \cite{MPSW,Grocholski:2019pqj}.
\paragraph*{Acknowledgements.}
\noindent
We acknowledge useful conversations with C\'edric Lorc\'e.   The work of J.W. is supported by the grant 2017/26/M/ST2/01074 of the National Science Center in Poland, whereas the work of L. S. is supported by the grant 2019/33/B/ST2/02588 of the National Science Center in Poland.  This project is also co-financed   by the Polish-French collaboration agreements  Polonium,  by the Polish National Agency for Academic Exchange and COPIN-IN2P3 and by the European Union's Horizon 2020 research and innovation programme under grant agreement No 824093.
\section*{Appendix}
\numberwithin{equation}{section}
\setcounter{equation}{0}
\renewcommand{\theequation}{A.\arabic{equation}}

We are considering the amplitude for the QCD  process contribution to the electroproduction of two photons:
\begin{equation}
e(k,\mu)+N(P_1,s_1)\rightarrow e(k',\mu)+N(P_2,s_2)+\gamma(k_1,\lambda_1)+\gamma(k_2,\lambda_2)
\end{equation}
assuming particular helicities and spins combination $\mu=\mu'=+\,; s_1=s_2=+,\, ;\lambda_1=+,\lambda_2=-$.

We neglect the electron mass, which ensures helicity conservation along the electron line. Following \cite{Kleiss:1985yh}, we write the massive spinor momenta as the sum of two lightlike momenta (here in the simplest case of $\Delta_T= 0$ ) :
\begin{equation}
P_1 = p_1 +p_2 ~~; ~~P_2 = p'_1 +p'_2 ~~;~~p_2 =(1+\xi)p ~~;~~p_1 =\frac{M^2}{1+\xi}n~~;~~p'_2 =(1-\xi)p ~~;~~p'_1 =\frac{M^2}{1-\xi}n\,,
\end{equation}
which allows to decompose the nucleon spinors as a linear combination of two massless fermion spinors:
\begin{equation}
U(P_1,+)=\frac{s(p_1,p_2)}{M}u_+(p_1)+u_-(p_2)~~;~~U(P_1,-)=\frac{t(p_1,p_2)}{M}u_-(p_1)+u_+(p_2)\,,
\end{equation}
with the $s(p_1,p_2) =-t^*(p_1,p_2)$ products defined as :
\begin{equation}
s(p_1,p_2) = (p_1^y+ip_1^z)\sqrt{\frac{p_2^0-p_2^x}{p_1^0-p_1^x}}-(p_2^y+ip_2^z)\sqrt{\frac{p_1^0-p_1^x}{p_2^0-p_2^x}}\,,
\end{equation}
so that $s(p_1,p_2) s^*(p_1,p_2) =t(p_1,p_2) t^*(p_1,p_2) = 2 p_1\cdot p_2$.
The photon polarization vectors read in the $p-$gauge:
\begin{equation}
\epsilon^\mu_\pm(k_i) =N_i\bar u_\pm(k_i) \gamma^\mu u_\pm(p)\quad N_i=\frac{1}{2\sqrt{k_i\cdot p}}\,.
\end{equation}
The amplitude is written as the product of the leptonic and hadronic parts
\begin{equation}
i\mathcal{M}=\frac{i}{Q^2}L^\alpha H_\alpha \,,
\end{equation}
where the leptonic part reads 
\begin{equation}
L^\alpha=\bar{u}_+(k')\gamma^\alpha u_+(k) \,,
\end{equation}
and the hadronic part reads, focusing on the contribution of the $H(x,\xi, t)$ GPD,
\begin{eqnarray}
&&
H_\alpha=\sum_q\frac{Q^3_q}{2}\int dx~\textrm{Tr}
\left[\tilde{\mathcal{M}}_\alpha(x)\not p\right]F^q_{++}(x)\,,
\end{eqnarray}
with $Q_u= 2/3, Q_d=-1/3, ...$ and
\begin{eqnarray}
&&F^q_{++}(x)=\frac{H^q(x)}{s}\bar{U}_+(P_2)\not n U_+(P_1)=\frac{H^q(x)}{s}\left[\frac{t(p'_2,p'_1)s(p_1,p_2)}{M^2}s(p'_1,n)t(n,p_1)+t(p'_2,n)s(n,p_2)\right]
\end{eqnarray}
\begin{equation}
i\mathcal{M}=i\frac{1}{Q^2}\frac{1}{2s}\sum_q Q_q^3\int dx\;  \textrm{Tr}\left[\tilde{\mathcal{M}}_\alpha(x)\not p\right]L^\alpha H^q(x)
\end{equation}
\begin{figure}
 \includegraphics[width=15cm]{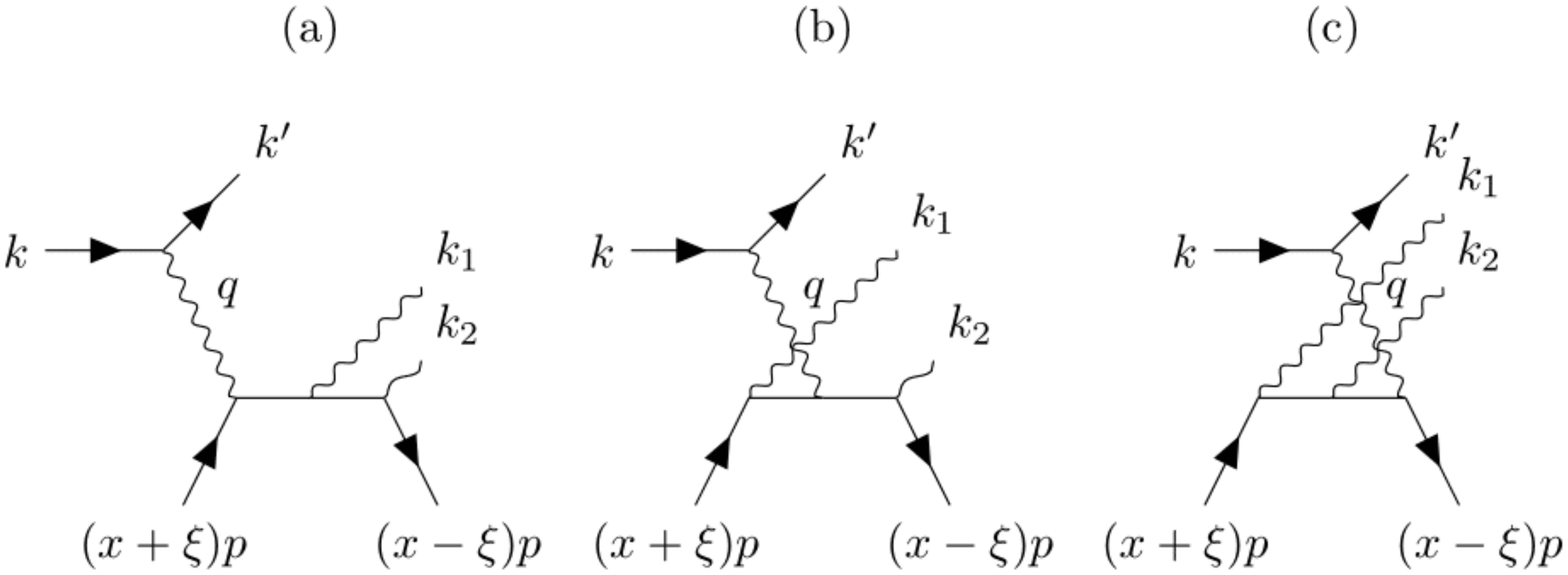}
\vspace{-2cm}\caption{The Feynman diagrams contributing to the QCD process; the other 3 diagrams are obtained by $k_1 \leftrightarrow k_2$ interchange. }
\label{FigApp}
\end{figure}
The three diagrams of Fig.\ref{FigApp} give the following contributions:
\begin{eqnarray}
&&\tilde{\mathcal{M}}^{(a)}_\alpha=\not\epsilon^*_-(k_2)\left[\frac{(x-\xi)\not{p}+\not k_2}{D_1(x)}\right]
\not \epsilon^*_+(k_1)\left[\frac{(x-\xi)\not p+\not k_1+\not k_2}{D_2(x)}\right]\gamma_\alpha\\
&&\tilde{\mathcal{M}}^{(b)}_\alpha=\not\epsilon^*_-(k_2)\left[\frac{(x-\xi)\not{p}+\not k_2}{D_1(x)}\right]
\gamma_\alpha\left[\frac{(x+\xi)\not p-\not k_1}{D_3(-x)}\right]\not\epsilon^*_+(k_1)\\
&&\tilde{\mathcal{M}}^{(c)}_\alpha=\gamma_\alpha\left[\frac{(x+\xi)\not p-\not k_1-\not k_2}{D_2(-x)}\right]\not\epsilon^*_-(k_2)
\left[\frac{(x+\xi)\not{p-}\not k_1}{D_3(-x)}\right]\not\epsilon^*_+(k_1) \,,
\end{eqnarray}
with the propagator denominators:
\begin{eqnarray}
D_1(x)&=&\alpha_2 s(x-\xi+i\varepsilon)\equiv\alpha_2 sD(x,\xi)\,,\nonumber \\
D_2(x)&=&ys(x+\xi-\frac{Q^2}{sy}+i\varepsilon)=ysD(x,-\xi')\,,\\
D_3(x)&=&\alpha_1s(x-\xi+i\varepsilon)=\alpha_1sD(x,\xi)\,,\nonumber
\end{eqnarray}
with $D(x,\xi)=x-\xi+i\varepsilon$ and where we denote $\xi'=\xi-\frac{Q^2}{sy}$.
Denoting
\begin{equation}
\mathcal{A}^{(x)}=\textrm{Tr}\left[\tilde{\mathcal{M}}^{(x)}_\alpha\not p\right]L^\alpha\qquad;\qquad
^c\mathcal{A}^{(x)}=\textrm{Tr}\left[\tilde{\mathcal{M}}^{(x)}_\alpha(k_1\leftrightarrow k_2)\not p\right]L^\alpha \,,
\end{equation}
we get:
\begin{eqnarray}
&&\mathcal{A}^{(b)}+^c\mathcal{A}^{(b)}=8N_1N_2 t(k,k_1)s(k_2,k')\left[\frac{1}{D(x,\xi)D(x,-\xi)^*}+(x\rightarrow -x)\right]\,; \nonumber \\
&&\mathcal{A}^{(a)}+^c\mathcal{A}^{(a)}+\mathcal{A}^{(c)}+^c\mathcal{A}^{(c)}=\frac{8N_1N_2}{ys}\nonumber\\
&&\left(\left(
s(p,k_2)t(k_1,p)t(k,p)s(p,k')+t(p,k_1)s(k_2,p)s(k',p)t(p,k)\right)\left[\frac{1}{D(x,-\xi')}+(x\rightarrow -x)\right]+\right.\\
&&\left.+\left(s(p,k_2)t(k_1,k_2)t(k,p)s(k_2,k')+t(p,k_1)s(k_2,k_1)s(k',p)t(k_1,k)\right)
\left[\frac{1}{D(x,\xi)D(x,-\xi')}+(x\rightarrow -x)\right]\right)\,.
\nonumber
\end{eqnarray}
Denoting the C-odd charge weighted GPD combination as:
\begin{eqnarray}
&&H^{(-)}(x)=\sum_q Q_q^3\left(H^q(x)+H^q(-x)\right)\,,
\end{eqnarray}
we need the following integrals
\begin{eqnarray}
&&I(\xi)=\int^1_{-1}\frac{H^{(-)}(x)}{x-\xi+i\varepsilon}dx=\int^1_{-1}\frac{H^{(-)}(x)-H^{(-)}(\xi)}{x-\xi}dx+H^{(-)}(\xi)\log\frac{1-\xi}{1+\xi}-i\pi H^{(-)}(\xi)\,,\\
&&{\sum_q Q_q^3\int^1_{-1}\left[\frac{1}{D(x,\xi)D(x,-\xi)^*}+(x\rightarrow -x)\right]H^q(x)dx=\frac{1}{2\xi}\left(I(\xi)-I^{*}(-\xi)\right)}\,,\\
&&{\sum_q Q_q^3\int^1_{-1}\left[\frac{1}{D(x,-\xi')}+(x\rightarrow -x)\right]H^q(x)dx=\frac{1}{2}\left(I(-\xi')-I^{*}(\xi')\right)}\,,\\
&&{\sum_q Q_q^3\int^1_{-1}\left[\frac{1}{D(x,\xi)D(x,-\xi')}+(x\rightarrow -x)\right]H^q(x)=\frac{1}{2(\xi+\xi')}\left(I(\xi)-I^q(-\xi')-I^{*}(-\xi)+I^{*}(\xi')\right)}\,,
\end{eqnarray}
that we evaluate numerically with the GPD parametrizations of Ref. \cite{GK}.

The resulting expression for the helicity amplitude is then given by: 
\begin{eqnarray}
&&\mathcal{M}=\frac{1}{Q^2}\frac{1}{2s}\left[\frac{t(p'_2,p'_1)s(p_1,p_2)}{M^2}s(p'_1,n)t(n,p_1)+t(p'_2,n)s(n,p_2)\right]
8N_1N_2 \nonumber \\
&&\left\{
t(k,k_1)s(k_2,k')\frac{1}{2\xi}\left[I(\xi)-I^*(-\xi)\right]+\right.\\
&&+\left(
s(p,k_2)t(k_1,p)t(k,p)s(p,k')+t(p,k_1)s(k_2,p)s(k',p)t(p,k)\right)\frac{1}{2sy}\left[I(-\xi')-I^{*}(\xi')\right]+\nonumber\\
&&+
\left(s(p,k_2)t(k_1,k_2)t(k,p)s(k_2,k')+t(p,k_1)s(k_2,k_1)s(k',p)t(k_1,k)\right) \nonumber \\
&&\left.
\frac{1}{2sy}\frac{1}{\xi+\xi'}\left[I(\xi)-I^q(-\xi')-I^{*}(-\xi)+I^{*}(\xi')\right]
\right\}\nonumber\,.
\end{eqnarray}

\end{document}